\documentclass[12pt,preprint]{aastex}

\def\dw#1{}			
\def\jm#1{}

\newcommand{\op}{Ly$\alpha$\ }

\begin{document}


\title{Small Scale Structure at High Redshift:
III. The Clumpiness of the Intergalactic Medium on Sub-kpc Scales \altaffilmark{1}\\
 }

\vskip 1.5cm

\author{Michael Rauch\altaffilmark{2}, Wallace L.W. Sargent\altaffilmark{3},
Thomas A. Barlow\altaffilmark{3}, Robert F. Carswell\altaffilmark{4}}
\altaffiltext{1}{The observations were made at the W.M. Keck Observatory
which is operated as a scientific partnership between the California
Institute of Technology and the University of California; it was made
possible by the generous support of the W.M. Keck Foundation.}
\altaffiltext{2}{Carnegie Observatories, 813 Santa Barbara Street,
Pasadena, CA 91101, USA}
\altaffiltext{3}{Astronomy Department, California Institute of Technology,
Pasadena, CA 91125, USA}
\altaffiltext{4}{Institute of Astronomy, Cambridge University, Madingley Road,
Cambridge CB30HA, UK}
\vfill

\begin{abstract}
Spectra obtained with the Keck HIRES instrument of the Lyman $\alpha$
forests in the lines of sight to the A and C components of the
gravitationally lensed QSO, Q1422+231, were used to investigate the
structure of the intergalactic medium at mean redshift $<z> \sim 3.3$
on sub-kpc scales.  We measured the cross-correlation amplitude between
the two Ly$\alpha$ forests for a mean transverse separation of $120
h_{50}^{-1}$pc, and computed the RMS column density and velocity
differences  between individual absorption systems seen in both lines
of sight.  The RMS differences between the velocity centroids of the
Lyman $\alpha$ forest lines were found to be less than about 400
ms$^{-1}$, for unsaturated  HI absorption lines with column densities
in the range $12<\log{\mathrm N(HI)}<14.13$.  The rate of energy
transfer into the low density IGM on a typical scale of 100 pc seems to
be lower by 3-4 orders of magnitude than the rate measured earlier for
strong CIV metal absorption systems.  The tight correlation between HI
column density and baryonic density in the intergalactic medium was
used to obtain a conservative upper limit on the RMS fluctuations of
the baryonic density field at $<z>=3.26$, namely,
$\sqrt{\left<\left(\Delta\log \rho\right)^2\right>} \leq
3.1\times10^{-2}$ on a scale of $110\ h_{50}^{-1}$pc. The fraction of
the absorption lines that are different across the lines of sight was
used to determine the filling factor of the universe for gas which has
suffered recent hydrodynamic disturbances. We thereby derived upper
limits on the filling factor of galactic outflows at high redshift.
Short-lived, short-range ancient winds are essentially unconstrained by
this method but strong winds blowing for a substantial fraction of a
Hubble time (at z = 3.3) appear to fill less than 20\% of the volume of
the universe.
\end{abstract}


\newpage

\clearpage

\section{Introduction}

Much astronomical effort has been
expended to investigate the structure of the universe on ever larger
scales. Surveys of galaxies in emission, studies of intergalactic gas
clouds in absorption, and searches for anisotropy in the cosmic
microwave background indicate a universe approaching homogeneity on scales
larger than a few hundred Mpc (e.g., Peebles 1993).  Velocity and
density variations, smoothed over increasingly larger volumes, decrease
steadily, in agreement with current cosmological n-body and
hydro-simulations. 

In this paper, we investigate how the density and velocity fluctuations 
behave on the {\em smallest} scales. Even
limiting ourselves to  randomly chosen regions (which have densities
far short of those encountered in stars and other highly collapsed
objects) the answer is complex and must depend strongly on the  local mean
density.  Large density fluctuations on very small (parsec)
scales occur in the {\it interstellar} medium. The {\it intergalactic} medium
(IGM), at least as observed at high redshifts, occupies an
intermediate range between large scale homogeneity and small scale
fragmentation. The combination of low density and photo-ionization heating by
the meta-galactic UV background conspire to prevent catastrophic
collapse and cooling. In the absence of discrete sources of momentum,
heat, or ionization, any small scale structure below the Jeans mass
should be smoothed out by the thermal pressure of the intergalactic gas.

The nature of small scale structure in the IGM has received renewed
interest with the advent of high resolution gasdynamic simulations of
structure formation. It is possible to measure various cosmological and
astrophysical quantities by comparing hydro-simulations of the general
baryon distribution with observations of \op forest absorption lines
(e.g., Cen et al.  1994; Petitjean, Muecket, \& Kates 1995; Zhang,
Anninos \& Norman 1995; Hernquist et al.  1996; Miralda-Escud\'e et
al.  1996; Wadsley \& Bond 1996; Zhang et al.  1997; Croft et al 1997;
Theuns et al.  1998; Bryan et al 1999; Dav\'e et al 1999).  In spite of
the successes of the models in reproducing the observed Ly$\alpha$
forest absorption, there is some concern that the finite mass
resolution may limit quantitative conclusions as to the thermodynamics
and ionization state of the gas.  For example, the optical depth of an
absorption line for a given value of the background density is
approximately proportional to the clumpiness of the gas.  Consequently,
a locally denser gas could, to some degree, mimic a universe with a a
larger baryon density and still satisfy the observational constraints
(see Rauch et al 1997; Weinberg et al. 1997).  Thus, estimates of the
baryon density in the intergalactic medium depend on either the
simulations properly resolving those low- to intermediate column
density structures in the IGM that contain most of the baryons, or on
the absence of any significant small scale structure. Since the Jeans
mass $M_J$ varies proportional to $\rho^{-1/2}$ ($\propto
(1+z)^{-3/2}$), there is always a redshift where the mimimum resolved
mass exceeds $M_J$, so small scale structure present at high redshift
could pose a problem.  There are good reasons for concern that such
structure may exist at some level.  Velocity gradients of several
kms$^{-1}$, or appreciable density differences over distances of a few
hundred parsecs in {\it low column density} gas, might imply that
already at high redshift much of the matter is trapped in unresolved,
small potential wells; alternatively, the gas may have been stirred by
non-gravitational agents such as galactic winds and explosions (e.g.,
Couchman \& Rees 1986). The widespread metal enrichment in the
intergalactic medium shows that such galactic outflows must have
occured. There have been several theoretical attempts trying to
understand this result (e.g., Ferrara, Pettini, \& Shchekinov 2000;
Theuns, Mo \& Schaye 2001,  ; Cen \& Bryan 2001; Madau, Ferrara, \&
Rees 2001; Aguirre et al 2001), but little is known about the
hydrodynamical consequences of these ancient winds for the IGM.
Finally, it has been argued that even at high redshift many Ly$\alpha$
forest absorption lines are formed in extended galactic halos (Chen et
al 2000), in which case one should expect to see small scale
hydro-dynamic disturbances in the gas caused by stellar feedback.

The only way to test for the possibility of hidden small scale structure
and of a universe grainier than envisaged by the simulations  is to
{\it measure} by how much the universal density and velocity fields
fluctuate on a given spatial scale.  Differences in the absorption
pattern of the Ly$\alpha$ forest in closely spaced lines of sight to
gravitationally lensed QSOs can be used to measure the clumpiness of the
IGM on sub-kpc scales, about an order of magnitude smaller than can
currently be resolved by cosmological simulations.  Such observations
have  been used to establish lower limits on the sizes of the
Ly$\alpha$ forest clouds (e.g., Young et al., 1981; Weymann \& Foltz
1983; Foltz et al.  1984; Smette et al.  1993,1995; Bechtold \& Yee
1995; Petry et al 1998).  Most of the above investigations have shown
the difference between the lines of sight to be small.  In the previous
paper (Rauch, Sargent, \& Barlow 2001, paper II), the second in a
series of papers on the small scale properties of various gaseous
environments at high redshift, we found that the metal enriched gas
observed in CIV absorption systems, apparently produced in extended regions 
around
redshift galaxies, shows signs of having been disturbed
hydrodynamically on time-scales similar to those relevant for
recurrent star-formation. This is consistent with the CIV gas clouds
being the result of ancient galactic outflows. The CIV absorbing gas is
mostly related to strong saturated Ly$\alpha$ forest systems, and the
question remained open as to how far the lower density IGM (causing
the much more frequent, unsaturated Ly$\alpha$ forest systems) has been
affected.  In the present paper, we use the Keck HIRES spectrograph
(Vogt et al.  1994) to investigate the fabric of the Ly$\alpha$ forest
proper on sub-kpc scales at  high signal-to-noise ratio and with a
velocity resolution sufficient to resolve the width of the Lyman lines
and detect velocity gradients of a few hundred ms$^{-1}$.

The most suitable object known for this sort of study is the
ultrabright, quadruple, gravitationally lensed QSO Q1422+231 ($z_{em} =
3.61$; Patnaik et al 1992). This object has been observed for similar
purposes previously at low resolution. Bechtold \& Yee (1995) concluded 
from
ground-based spectra that there are differences of typically
about 50\%\ in HI column density for beam separations on the order of
100 parsec.  In contrast, Petry, Impey \& Foltz (1998), from HST FOS
spectra found very little difference both in column density and
velocity between the various lines of sight. In section 2 we describe
Keck HIRES (Vogt et al. 1994) spectra of the A and C images of
Q1422+231. The properties of this lens system are summarized on the
CASTLES web site (http://cfa-www.harvard.edu/castles/B1422.html). The
observations of images A and C of Q1422+231 and the methods used to
reduce the data are discussed in section 2. Section 3 contains an
account of the measurement of the differences in the Lyman $\alpha$
lines in the spectra of the two images. Our conclusions are summarized
in section 4.

\section{Observations and data analysis}

Q1422+231 consists of four images. Images A (g= 16.92 magn.), B (16.77 magn.)
and C (17.44 magn.) are almost along a line with C separated by 1.64"
from A and 0.76" from B. Image D (20.56 magn.) is well separated from the rest.
We chose to obtain HIRES spectra of A and C, using a 0.56" slit and only
observing in conditions of excellent seeing. Much of the guiding was done 
manually and the position angle of the slit was chosen so as to keep image 
B out of the slit. In some exposures we guided on the outer edge of the image
in order to ensure minimal contamination from B. 
The dataset is the same as described in papers I (Rauch, Sargent \& Barlow 1999)
and II (Rauch, Sargent \& Barlow 2001).
The spectra were reduced as described in Barlow \& Sargent (1996).
The individual A and C exposures were matched in flux to a
template produced by adding up all A spectra, using polynomial fits
typically of 8th order per echelle order.  Care was taken to include
only points with flux levels close to the continuum in the fitting
regions, and to avoid absorption lines.  This way, the large scale
features in the spectra become identical (the information on large
scale differences between the QSO continua is lost) but features on scales 
up to $\sim$ 300-400 kms$^{-1}$ are retained.
Thus, differences among individual absorption line profiles up to that velocity 
scale survive
our matching procedure and remain unaffected. The resulting spectra
were normalized to a unit continuum using spline functions.
Voigt profiles were fitted to the absorption lines 
using the fitting routine VPFIT (Carswell et al. 1991, see 
http://www.ast.cam.ac.uk/~rfc/vpfit.html), until the fit exceeded 
a minimum probability
(1\%) to produce by chance a $\chi^2$ as high as the one attained.  We
departed here from previous such analyses in that the whole spectrum
was fitted (in continuous pieces), without  selecting
significant absorption features by eye.  We thus do not have to assign
significance thresholds to individual lines, a procedure
which is most meaningful for single unblended, strong absorption
components, but which becomes uncertain in more complicated situations. 
Another problem,
peculiar to relatively high signal-to-noise ratio spectra such as those 
obtained with HIRES, is that it is hard to obtain
satisfactory fits merely by superimposing Voigt profiles, even if Voigt
profiles are a good physical model.  When S/N ratios exceed a
few tens, the errors in the fit are invariably dominated by systematic
uncertainties in the placement of the continuum level. Therefore, it was 
decided to treat the continuum in every fitting region
(average length 28 \AA ) as a free parameter. This resulted in a mean
drop of the continuum level over the region [4747, 5630] \AA\
(i.e., between Ly$\beta$ and Ly$\alpha$ emission, where the \op analysis 
in the present paper was performed) to 99.3 \%
of the original level, with a standard deviation of 2.7\% of the
original continuum level, i.e., the total continuum level remained
virtually unaffected by introducing locally free continua, but the
local continua were adjusted typically by 2.7\% up or down, to obtain an
optimal fit. 

Since the A and C images are so close to B we have to be concerned that
blending of the spectrum of an image with light from the other images
might have reduced the differences seen between the spectra. However,
our spectra  show strong column density differences between the A and C
images in some low ionization metal lines in the Lyman $\alpha$ forest
(see paper I) which would have been significantly reduced if spillover
of light from different images had been a serious problem\footnote{we
have also obtained as yet unpublished spectra of the lensed QSO UM673
with HIRES. In this case the images A and B have a separation of 2.2"
so there is little risk of mixing the light from different images. The
spectra have significantly lower signal-to-noise ratio than the data on
Q1422+231 discussed here, but the absence of differences between the
low column density Ly$\alpha$ forest lines in UM673A,B is consistent
within the errors with the very small differences found in the present
paper.}. Our technique of matching the two spectra to a common template
as described earlier {\it does} take out the large scale fluctuations
but not those among individual lines. Again, the differences seen in
metal line systems suggest that this is not a problem. Finally, while
they may affect the measured {\it column densities}, none of these
effects could conspire to reduce any actual intrinsic {\it velocity}
differences between the lines of sight to values as tiny as those found
here.

The lowest redshift covered by the selected spectral region, $z=2.905$,
corresponds to a transverse separation $\Delta r \sim 270$ pc.
Throughout the paper beam separations are computed for $H_0 = 50
$ kms$^{-1}$Mpc$^{-1}$ and $q_0=0.5$. The redshift of the lens is
taken to be z = 0.338 (Kundic et al. 1997).

\section{Differences between the lines of sight in the \op forest region}

Here we describe three main approaches to measuring differences
between the \op forests in the two adjacent lines of sight:
a global cross-correlation analysis; a search for column density
and velocity differences among individual absorption lines; 
and a pixel by pixel measurement of relative fluctuations in the optical depth
between the lines of sight.

\subsection{Global correlation analysis}

We can study global differences in the \op forest region
by measuring the cross-correlation function, $\xi_{cc}$, over the total,
useable length of both spectra.
We define this quantity by 

\begin{eqnarray}
\xi_{cc} (\Delta v, \Delta r) \equiv
{ < (F_{\bf r}(v) - < F_{\bf r} > ) \cdot
    (F_{{\bf r}+\Delta {\bf r}}(v + \Delta v) - < F_{{\bf r}+\Delta {\bf r}} > ) > \over
   \sqrt{< (F_{\bf r}(v) - < F_{\bf r} >)^2 > \cdot
   <(F_{{\bf r}+\Delta {\bf r}}(v + \Delta v)-<F_{{\bf r}+\Delta {\bf r}}>)^2>} } ~. \label{eq:ccf} 
\end{eqnarray}

Here $F_{\bf r}$ and $F_{{\bf r}+\Delta {\bf r}}$ are the pixel flux
values of the two spectra, separated by $\Delta r$ on the plane of the
sky. The velocity coordinate along the line of sight is $v$ (where $dv
= c d\lambda/\lambda$), and $\Delta v$ is the velocity lag. The averages
are taken over the whole velocity extent of the spectrum. For $\Delta
r$ = 0 we get the usual auto-correlation function $\xi_{cc}(\Delta v,
0)$, while for $\Delta v$ = 0 we have the cross-correlation as a
function of transverse separation only.  The function
is defined so as to satisfy $\xi_{cc} (0,0)=1$. With large scale
velocity correlations ($>$ 1000 kms$^{-1}$) expected to be absent or
weak (Sargent et al. 1980), the autocorrelation function (on scales
$\sim$ 100 kms$^{-1}$) mostly measures the \op line width and the weak
small scale clustering of \op forest systems (e.g., Webb 1987, Rauch et
al.  1992).  We apply the correlation analysis to most of the spectral
region between Lyman $\beta$ and Lyman $\alpha$ emission, [4737, 5630]
\AA . Only regions [4747, 4875], [4895,4997], [5003,5221], [5227,5409],
[5419,5473], and [5480,5630] \AA\ were used in the analysis in order to 
avoid contamination by
known metal systems.  These regions are shown in fig. 1; the hatched
parts were excluded.  The resulting  mean redshift $\overline{z}$ =
3.26036 of the remaining sample corresponds to a mean beam separation
$\overline{\Delta r}$ (=119 h$^{-1}_{50}$pc).  A small portion of the
spectrum is shown in fig. 2. The spectrum of image C (dotted line) is 
overplotted on that of A
(solid line), to illustrate how closely the spectra resemble each
other.

The function 
$\xi_{cc} (\Delta v,\overline{\Delta r})$ is shown in
fig. 3.  In particular, we obtain the "zero-lag"
cross-correlation function:
\begin{eqnarray}
\xi_{cc}(\Delta v = 0;\ \overline{\Delta r} = 
119\ h_{50}^{-1}\ {\rm pc})=\ 99.48\ \% .
\end{eqnarray}
 Thus, over a mean transverse distance of 120 pc
the appearance of the \op forest changes very little, and even the small
difference observed may
be due to noise or systematic errors.

\subsection{Comparison between individual absorption lines}

We next investigate how properties of individual absorption systems change with
the
separation on the plane of the sky.  We can quantify differences $\Delta x =
x_1 - x_2$ in any physical property $x$ of absorption systems common to
both lines of sight 1 and 2 by means of the distribution function of
$\Delta x$. (For example, consider the difference between the redshifts
of the lines, $z_1 - z_2$). In an isotropic universe, the mean, 
$\overline{\Delta x}$, vanishes, so the information is contained in the
shape of the distribution of differences, as characterised, for
example, by the {\it observed} scatter of the distribution,
$\sigma(\Delta x)$. The width of the {\it intrinsic} distribution,
$\sigma_{int}(\Delta x)$, differs from the {\it observed} one due to the
combined measurement errors, $\sigma(x_1)$ and $\sigma(x_2)$, and any 
systematic error $\sigma_{sys}$ resulting from the
occasional erroneous pairing of two unrelated absorption components.
The quantity $\sigma_{<}(\Delta x)$ = $\sqrt{\sigma(\Delta x)^2 -
\sigma(x_1)^2 - \sigma(x_2)^2}$ is then a strict {\em upper limit} on
the intrinsic scatter, $\sigma_{int}(\Delta x)$.

Here we concentrate on the distribution of {\em velocity} and {\em column 
density}
differences, $\Delta v$ and $\Delta \log N$. Inspection by eye shows that 
there are
only minute difference between the spectra of the A and C images.
Accordingly, it is legitimate to associate each Lyman $\alpha$ forest line 
in the spectrum of component A with a corresponding line in C.
Thus, after profile-fitting the spectra, we suppose
that an absorption component in spectrum C belongs to the same
underlying physical structure as a line in A if its redshift and column density
fall within a given redshift and column density window centered on a
similar component in the A spectrum.

We make the assignments totally automatically in
order to preclude human prejudice. In order to avoid too many random
associations we
restrict the allowed range of column density differences when measuring
the relative velocity distribution, and restrict the allowed
range of velocity differences when comparing the column densities.
Specifically, the redshifts had to agree to within $\pm$ 30 kms$^{-1}$
(there is no real prejudice as this is much larger than the actual
scatter), while column density differences $|$logN$_A$ -- logN$_C|$ $<$
1 were deemed acceptable. In order to avoid including ill-constrained
measurements and spurious cross-identifications, all lines included
were required to have relative Doppler parameter errors,  $\sigma(b)/b$
$<$3 and column density errors $\sigma(logN)/logN$$<$3. Doppler
parameter values had to fall in the interval [5, 300] kms$^{-1}$ to
exclude spuriously low values due to noise spikes and high
values due to unresolved blends. All error
estimates given below are 1-sigma deviations derived from the output of
the line profile fitting program.  Finally, to get rid of obvious
mis-associations, one iteration of a 4-$\sigma$ clipping
algorithm was performed. Column density differences larger than
4 standard deviations  of the mean of the distribution of column density
differences where rejected.  The analysis was limited to column
densities 12$<\log N<14.13$, the lower bound being close to the
detection limit, whereas the upper limit (14.13) corresponds to a
residual flux of about 2.4 \% (for a Doppler parameter of 28
kms$^{-1}$, typical of an unsaturated $z\sim 3$ Ly$\alpha$ forest line), a 
value just at the top of the linear part of the curve
of growth, but far enough away from  saturation to permit us to
see significant differences between the central optical depths.  This
column density regime is most relevant for the study of gravitational
structure formation since the selected absorption lines probe gas
overdensities still close to the linear regime of gravitational
instability  ($\Delta \rho/\rho\leq 15$) on the scale of individual
galaxies (e.g., Miralda-Escud\'e et al.
1996). 

\subsubsection{velocity differences}

Fig. 4 shows the observed histogram of velocity differences, $\Delta v
= v_A - v_C$\footnote{The original distribution did not have a zero
mean but showed an mean shift of spectrum C by 0.96 kms$^{-1}$. This
shift amounts to slightly less than half a pixel or a quarter of the
slit width. Its size and the circumstances of the observations make it
most likely that it was caused by systematic variations in the position
of the target on the slit: the targets were only observed in
exceptional seeing to minimize  contamination of the desired image by
the light from other images. Consequently, the seeing may often have
been better than the 0.574" slit width, leading to a dependence of the
wavelength zero point on the actual position of the target in the
slit.  This uncertainty may have been turned into a systematic shift by
our attempts to avoid spillover.  For the analysis below and for all the
plots from Fig. 4 on, the above shift was subtracted from the velocities of
the C spectrum.}.  The top plot of fig. 5 shows the observed standard
deviations of the distribution of velocity differences between the two
lines of sight (solid dots), versus the largest measurement error in
redshift for lines remaining in the sample. The open diamonds show the
predicted width of the distribution of velocity differences expected
from statistical measurement errors alone. It was computed as the
quadratic sum of the $1-\sigma$ error estimates produced by the Voigt
profile fitting routine. To search for intrinsic scatter we discard
those velocity differences  with large errors.  As the most uncertain
points are eliminated, the observed width of the distribution should
decrease, and any intrinsic scatter should emerge. We see that, the further 
one goes to the 
left along the abscissa, the line pairs remaining in the sample have
smaller and smaller maximum (and mean) redshift errors and the more
well defined are the measurements.  In other words, if the bulk
of the scatter is caused by measurement error, we should expect to see
the observed total scatter $\sigma(\Delta v)$ decreasing in step with,
and closely tracking, the measurement errors as increasingly stringent
error limits are applied and more and more data points are discarded
from the distribution. Aside from a certain scatter (probably caused by
mis-matched components) this behaviour is indeed observed in fig.5.
(The discrepancy at large values of the abscissa between the width of
the predicted distribution based on measurement errors alone, and the
observed distribution, is an artifact of the range of parameters we
imposed earlier. It is caused by our admitting only velocity
differences $< 30$ kms$^{-1}$ into the sample.  This is perfectly
justified given the range of the large majority of values in figure 4,
but it does get rid of some large outliers which result from automatic
mis-identifications of line pairs.)

To demonstrate the sensitivity of this method, we plot (for
illustrative purposes) the observed $\sigma(\Delta v)$ a second time
(dashed line in the top panel of fig. 5), but now with an additional
random, Gaussian velocity jitter of $\sigma$=1.5 kms$^{-1}$ imposed on
each line pair, i.e., the distribution of pair velocity differences has
now a width of $\sqrt{\sigma^2(\Delta v) + (1.5\ {\mathrm
kms}^{-1})^2}$.  On the left half of the diagram this jitter clearly
dominates the observed width, demonstrating that any intrinsic velocity
differences of that size would have been seen easily and that any
actual differences must be considerably less than 1.5 kms$^{-1}$.

There are 259 line pairs with individual redshift errors $\sigma(z)$$<$
4$\times$10$^{-5}$ with an {\it observed} total $\sigma({\Delta v})$
=1.7 kms$^{-1}$, and still 83 pairs in the bin with the smallest
errors, $\sigma(z)$$< 1\times10^{-5}$ ($\sigma({\Delta v})$ =0.6
kms$^{-1}$).  The upper limits $\sigma_{<}(\Delta v)$ on the {\it
intrinsic} $\sigma_{int}(\Delta v)$ (where the measurement errors have
been subtracted in quadrature from the observed width) are formally 0.4
and 0.2 kms$^{-1}$, respectively. Given the scatter between the
measurement errors and total $\sigma(\Delta v)$ visible in the plot,
the intrinsic distribution is probably consistent with zero width.  We
can compare these values with the  velocity differences expected from
the Hubble expansion alone. If Ly$\alpha$ clouds expand with the Hubble
flow the projected velocity differences between two lines of sight,
(modelling the geometry by an expanding slab intersecting both lines of
sight at an angle $\alpha$), would be on the order of $\Delta v_H/\tan\alpha =  H(<z>)
\ r \sim 50\ {\mathrm ms}^{-1}$, where $ H(<z>=3.26)= 440 h_{50}$
kms$^{-1}$Mpc$^{-1}$ was adopted as the Hubble constant at the mean
observed redshift.  This is well below our detection threshold and our
null result is consistent with the clouds following the Hubble flow.

It is tempting to try using the measurement of $\sigma(\Delta v)$ to get a rough idea of the
energy deposited in the gas in the form of turbulence. In paper II we had applied
the Kolmogorov relation between the structure function $B(r)$, and the 
energy input rate $\epsilon$ at scale $r$,
\begin{eqnarray}
B(r) \approx \overline{[v_1(r') - v_2(r'')]^2}\  \approx 
\ (\epsilon\ r)^{2/3}
\end{eqnarray} 
(e.g., Kaplan \& Pikelner 1970;  here $v_1(r')$ and $v_2(r'')$ are the
velocities along the line of sight, and $r$ is the spatial separation
between the lines of sight, and the average is taken over all
absorption line pairs with mean transverse separation $r$) to get an
approximate estimate of the energy input rate $\epsilon$ in CIV
absorbing clouds. These gas clouds correspond to strong, usually
saturated Ly$\alpha$ forest systems and overdensities higher than a few
tens.  This approach could be justified then as we had indeed observed
turbulent energy on a wide range of scales and had found a rise of
$B(r)$ with $r$; moreover the relevant time scales for energy
dissipation were short, so turbulent energy  must have been propagating
down over a range of scales close to the epoch of observation.  Now we
are applying this scheme to the lower density gas giving rise to
unsaturated low column density Ly$\alpha$ lines, which has densities
within a factor of 10 -- 15 of the mean density of the universe, and,
at average, must be situated much further away from sources of stellar
energy. Here it is more likely that any energy injection by stellar
feedback is intermittent or happened only once long ago so a steady
state (one of the conditions for the validity of the Kolmogorov law) may not
exist (or no longer exists).

With this warning in mind we proceed to calculate $\epsilon$ for our
present low column density Ly$\alpha$ forest sample. The quantity
$\epsilon$ is still a measure of the energy flow on a given scale, but
is not necessarily related to the current primary energy input on
larger scales (e.g., by explosions or winds). If we identify our measured
variance $\sigma^2(\Delta v)$ with $B(r)$, we can solve for the energy
transfer rate\footnote{Because of the uncertain geometry of the clouds and
our taking the known projected transverse separation instead of the
(unknown) actual 3-dimensional separation between points (see the
discussion in paper II) this approach can only provide a rough, order
of magnitude estimate of $\epsilon$.} at scale $r$:

\begin{eqnarray}
\epsilon\sim \frac{\sigma^3(\Delta v)}{r}.
\end{eqnarray} 

Adopting the above value $\sigma_{int}(\Delta v)=0.4\ $kms$^{-1}$, and 
$\overline{r} = 110 h_{50}^{-1}$ pc, 
\begin{eqnarray}
\epsilon\sim 2\times 10^{-7} h_{50}\ {\rm erg\ g}^{-1} {\rm s}^{-1}.
\end{eqnarray}
Thus the intergalactic medium giving rise to the unsaturated parts (i.e., most)
of the Ly$\alpha$ forest at z=3.26 is experiencing energy transfer at a rate three to 
four orders of magnitude less than the gas giving rise to the
CIV absorption systems dealt with in paper II\footnote{but note that the
CIV turbulence was measured on a three times larger scale, 300 pc, so the 
discrepancy may be less.}, and six to seven orders of magnitude
less than the values in the Orion nebula, for example.

\subsubsection{column density differences}

A scatter plot of the column densities in A and C is given in fig. 6. The
scatter is larger in the C image because of the lower signal-to-noise
ratio there.  Some of the outliers may come from the ambiguity of the Voigt 
profile models (which occasionally may lead to similar absorption complexes 
being modelled automatically by two rather
different sets of Voigt profile components), while others may be genuine
differences due to unidentified metal lines.
The distribution of column density differences is shown in
fig.7. Two components were considered to correspond to each other, when
they were nearest neighbors, lay within 30 kms$^{-1}$ of each other,
and had column densities differing by less than 1.0 dex.  Fig. 8  shows the scatter versus measurement error
diagram for column densities, similar to Figure 5.  For the best
defined samples (those with measurement errors around
5$\times$10$^{-2}$ in $\log N$) the corrected upper limit to the
intrinsic scatter is $\sigma_{<}(\Delta \log N)$$\sim$
$2\times10^{-2}$. We note that it makes sense to discuss an
`optimum sample' because discarding those data points with the
largest measurement error leaves better measurements but also
reduces the sample size and thus the statistical significance.
Initially throwing away the points with the worst error will improve
the precision with which sample mean and variance are known, but the
precision of these statistical values decreases with the square root of
the number of data points, $\sqrt{\cal{N}}$ and if enough points have
been lost the penalty gets larger than the gain. In the appendix we
have defined a function of merit, $E$, which is 
positive as long as it
is beneficial to discard data points, and crosses zero towards
negative values if the sample becomes to small. The bottom panels of
figures 5 and  8 show this function.  For example, from the column
density case (fig. 8) we find that the sample has
been cut down to optimal size by
retaining only data points with a maximum column density error $\leq
0.05$.

Most interesting from an  astrophysical point of view is whether we can 
put limits on the total baryon density fluctuations on the scale
of the transverse separations of the two lines of sight.
The relation between 
the HI column density N(HI) and the baryon density for Jeans-smoothed
IGM regions can be written 
as $$ N(HI)\propto \rho_{bar}^{\alpha}$$ with $\alpha$ = 1.37 -- 1.5 (Schaye 2001). For this 
power law the distribution
of {\it column density} differences $\Delta \log N$ (= $\log N_A$ -- $\log N_B$)  can be transformed
by simple scaling with 1/$\alpha$ = 0.73 (we use the smaller value  $\alpha=1.37$
to establish the most conservative upper limit) into a distribution of
{\it density} differences,
and the variance of the logarithmic baryon density becomes
\begin{eqnarray}
\left<\left({\Delta \log \rho}\right)^2\right> \leq \alpha^{-2} 
\left<\left(\log N_A - \log N_B\right)^2\right>
\end{eqnarray}
Adopting (conservatively) the pair sample with mean column density 
measurement errors  below
0.05 dex (153 lines; see fig. 8) 
we arrive at a value of
$2.5\times 10^{-2}$ for the corrected "intrinsic" scatter
and $4.3\times 10^{-2}$ for the uncorrected, observed scatter in 
$\Delta \log N$.
As the figures show, other choices of sample size give similar values.
Then we get  
\begin{eqnarray}
\sqrt{\left<\left(\Delta\log \rho\right)^2\right>} \leq 1.8\times10^{-2}
({\mathrm or}\ 3.1\times10^{-2}, {\rm uncorrected}), \end{eqnarray}
for the typical logarithmic change in density on a scale of 0.110
h$_{50}^{-1}$ kpc, i.e., the RMS fluctuation in the baryon density
is less than about 3.1 percent. From the close agreement between the width 
of the observed
distribution and the expected width from  measurement errors alone,
it is clear that the fluctuations are dominated by the measurement
errors and that the intrinsic fluctuations are consistent with being 
zero at our level of precision.

\subsection{The filling factor of 'disturbed' regions in the universe}

Here we address the question as to how much of the universe (by volume)
is hydrodynamically disturbed, for example by supernova
explosions or galactic winds along the two lines of sight. 
We consider a region in space at a given
redshift $z$ as 'disturbed', if the optical depth $\tau$ for HI Ly$\alpha$
absorption differs by more than a certain amount between the two lines
of sight. 

This is easy to measure unless the lines are saturated.  Many of the
saturated systems are associated with multi-component metal absorption
systems.  These systems exhibit substantial variations in the
absorption pattern across multiple, close lines of sight (e.g. Lopez et
al 1998; paper II), but the component structure and any differences in
the column densities are almost always completely obscured in the
strong, saturated HI Ly$\alpha$ absorption blends. Thus, to get a
conservative upper limit on the fraction of disturbed regions we assume
that saturated regions in the Ly$\alpha$ forest are always disturbed,
i.e., we treat them as if they had infinite optical depth differences.

Fig. 9 shows the distribution of the difference in  optical depth
$\Delta \tau = \mid\tau_A - \tau_C\mid$ between the two lines of sight
$A$ and $C$. The four different histograms show the results for various
column density ranges\footnote{the column densities quoted here are
calculated by treating the observed optical depths  as if they were
central optical depths of Gaussian lines with Doppler parameter 28
kms$^{-1}$.} (the column density in spectrum A was used to determine
whether a pixel pair fell into a certain range): the solid line is for
the whole spectrum, the dashed one for weak lines with $12.0 < \log N <
13.0$, the dash-dotted one for $13.0 < \log N < 14.0$, and the dotted
one for stronger, but still unsaturated, lines.  (A column density
$\log N = 14.13$ corresponds to a residual flux of 2.4\% for the center
of an absorption line with a typical Doppler parameter $b=28$ kms$^{-1}$).  In
the first histogram (solid line), saturated pixels (for practical
purposes those with $\log N > 14.13$), did not provide any information
and were assumed to have infinite optical depth differences $\Delta
\tau$; thus they do not appear in this histogram in the lower bins. The
relative proportions of the spectra in those four column ranges was
100\%, 36.6\%, 36.4\%, and 17.4\%, respectively.  To reduce
contributions to the optical depth differences from noise, both spectra
were smoothed with a box filter on the scale of typical line
broadening (28 kms$^{-1}$).

We found that 76.8\%, 98.7\%, 74.7\%, and 46.7\%, respectively, of all
the pixels in the four column bins had their optical depths  differ
between the two lines of sight by less than $\Delta \tau = 0.05$ (i.e.,
they contribute only to the first bin of fig. 9).  It is remarkable
that of the 36.6\% of all pixels which are in the column density range
$12.0 < \log N < 13.0$,\ 98.7\% agree to within $\tau < 0.05$.
Apparently, the low column density Ly$\alpha$ forest is virtually
undisturbed by any effects capable of producing differences in density
over a couple of hundred parsecs. The differences  in optical depth
increase with increasing column density as one would expect if higher
column density gas is more closely associated with galaxies, which are
the likely origin of any hydrodynamic disturbances.

The overall fraction of the Ly$\alpha$ forest with $\Delta \tau < 0.05$
irrespective of column density is $ 1- f_{LoS} = 76.8$\%. Since we have
counted regions close to or beyond saturation as having infinite
$\Delta \tau$ (see above),  $f_{LoS}$ is a strict upper limit on the
line-of-sight filling factor of the Ly$\alpha$ forest with 
absorption from processed gas capable
of producing a disturbance of that magnitude. 

\subsubsection{an upper limit to the filling factor of galactic wind bubbles}

The upper limit $f_{LoS} \leq 0.23$ on the fraction of disturbed pixels
in the Ly$\alpha$ forest can be used in a crude way to
estimate the number density and volume filling factor of galactic
outflows from their effect on the density field of the IGM, in sofar as
such variations appear as differences between the HI column densities
(Ly$\alpha$ forest opacities) between two close lines of sight. The
principle will be illustrated here for a model of windbubbles expanding into the
intergalactic medium (fig. 10).

We assume that dense shells of swept up material occasionally cross the
double lines of sight to a lensed background QSO. It can be shown (see
Appendix B) that even under unfavorable conditions  the shells should
have HI column densities high enough to be seen in absorption in a high
signal to noise ratio QSO spectrum like the one of Q1422+231 discussed here
(N(HI)$ > 10^{12}$cm$^{-2}$). The passage of the shell is very likely
to disturb the resemblance between the absorption lines in the two lines
of sight (Fig. 10). Ideally, the shell should produce two absorption
systems at its two points of intersection with the lines of sight. We
assume that there will be measurable differences in the absorption
lines between the lines of sight as long as new shell material keeps
crossing the lines of sight faster than the differences can be smoothed
out by pressure waves travelling at the sound speed.  Or in other
words,  a shell is no longer detectable for when the speed of
sound  $c$ exceeds the transverse velocity of the shell across the
lines of sight:
\begin{eqnarray}
\frac{\dot{R}}{R} b < c,
\end{eqnarray} 
where $b$ is the impact parameter of the lines of sight relative to the
center of the shell, and $R$ and $\dot{R}$ are the radius and expanding
velocity of the shell.

To convert the line-of-sight filling factor into a volume filling
factor a particular model for the disturbance and its observational
signature is needed. For illustrative purposes we adopt the supershell
model of MacLow \& McCray (1988), i.e., we assume that the
disturbances are wind-blown shells escaping from high redshift
galaxies.  For a given line-of-sight filling factor, the radius $R$,
velocity of the shell $\dot{R}$, volume filling factor $f_v$, and the
space density $n_{gal}$ of galaxies with a wind escaping into the IGM
can be written in terms of the mechanical luminosity $L_{38}$ (in units
of $10^{38}$ ergs/s), the density of the ambient medium $n_{-5}$ (in
units of $10^{-5}$ cm$^{-3}$), and the age of the wind bubble $t_7$ (in
units of $10^7$ years).  These relations are given in Appendix $B$.

Figures 11 and 12 show the dependence of the upper limits on  $n_{gal}$
and $f_v$ for winds of various 'strengths' (i.e., with the ratio
$L_{38}/n_{-5}$ as the distinguishing parameter) corresponding to the
upper limit $f_{LoS} \leq 0.23$ on the fraction of disturbed pixels in
the Ly$\alpha$ forest.

The model is highly oversimplified in that it assumes spherical bubbles
with a single age and a single combination of L$_{38}$ and $n_{gal}$,
propagating through a homogeneous medium (and the expansion of the
universe is not taken into account), but it contains several
features likely to be relevant to more realistic future models (figs.
11 and 12):

\smallskip

\noindent 1. The number of galaxies with winds originating close to the
epoch of observation ($z=3.26$) is essentially unconstrained by an
observation of the line of sight filling factor (fig.11).  This is
because the shell radii are so small that even a large number of
galaxies with winds would produce a negligible total absorption cross
section.  \smallskip

\noindent 2. Weak winds (low luminosity and / or large density of
ambient medium, i.e., $L_{38}/n_{-5} \leq 1$), because of their limited
lifetime, escape detection if they originated early (e.g., before
$z_{SF} \sim 4$ for a $L_{38}/n_{-5} \sim 0.1$ shell, or  before
$z_{SF} \sim 7$ for a $L_{38}/n_{-5} \sim 1$ shell; fig. 11).
\smallskip

\noindent 3. Strong winds ($L_{38}/n_{-5} > 10$) remain detectable even
if they started a Hubble time ago (fig. 11), and the upper limit on their
numbers becomes stronger with increasing redshift $z_{SF}$.

\noindent 4. The volume filling factor (fig. 12) is most
tightly constrained for recent rather than old winds, independent of
strength.

\noindent 5. Our observation cannot exclude that the universe is filled
with ancient, weak wind bubbles because they would no longer be
detectable with the present method by redshift $z \sim 3$ (see item 2 above).
However, since these weak winds do not reach far out from galaxies and
galaxies are not distributed randomly it is very unlikely that such
bubbles could fill a substantial fraction of the voids.  Strong winds
($L_{38}/n_{-5} > 10$) are much better constrained because of their
long lifetimes: if winds of strength $L_{38}/n_{-5} \sim 1000$ started
as early as $z_{SF} \sim 10$, they could be filling up to about 40\% of
the volume of the universe by redshift z = 3.26. If such strong winds
arose at $z_{sf} \sim 4$ they may fill up to 18\% .

\section{Conclusions}

We have applied various methods to search for and measure density and
velocity structure on 100 pc scales in the intergalactic medium at
$<z>=3.26$.  The smallness of the differences in column density and
velocity between the lines of sight confirm and extend the results of
Smette et al (1992,1995), and Petry, Impey \& Foltz (1998).  In the
unsaturated regions of the Ly$\alpha$ forest RMS differences between
the projected line of sight velocities over 120 pc do not exceed a few
hundred meters per second.  The RMS total density differences are
inferred to be less that 3\%, and are basically below the detection
limit. Visible differences between the lines of sight do occur
occasionally but are localized and  seem to be mostly limited to high
column density gas. Strong differences are known to occur from studies
of the higher column density metal absorption lines (papers I and II),
but these differences rarely show up in the Ly$\alpha$ region proper
because of saturation. Our results appear broadly consistent with
predictions by Theuns, Mo \& Schaye (2001) who model the impact of
stellar feedback on the appearance of the Ly$\alpha$ forest and find
that stellar feedback will mostly affect the higher column density,
Lyman Limit absorption systems.  The suggestion (Chen et al 2000) that
most absorption lines with a rest frame equivalent width $>0.32 \AA\ $
(corresponding to N(HI)$\sim 6\times 10^{13}$ cm $^{-2}$ for a Doppler
parameter $b = 28$kms$^{-1}$) are related to galactic halos (where we
might expect to see hydrodynamic disturbances) is not contradicted by
our results because close to 90\% of the Ly$\alpha$ forest lines above
our chosen column density threshold ($10^{12}$ cm$^{-2}$) have lower
column densities.

Our results are consistent with unsaturated Ly$\alpha$ systems being
mostly featureless at the few percent level on scales smaller by an
order of magnitude than state-of-the art cosmological hydro-simulations
can currently resolve.  Apparently, the finite resolution of the
simulations does not lead to a significant underestimate of the
clumpiness of the baryon distribution, for densities typical of the
unsaturated Ly$\alpha$ forest, i.e., overdensities up to a factor
10-15. However, our observations only provide a snapshot of the
properties of the Ly$\alpha$ forest at $z\sim 3.3$ and cannot exclude
the existence of earlier non-gravitational processes in the IGM if, by
the time of observation, their hydrodynamical traces have been
obliterated (see also below).

Notwithstanding doubts about the applicability  of the Kolmogorov law
to the low density IGM, we have tentatively derived the rate of
turbulent energy input into the low density IGM. The turbulent energy
in the unsaturated low column density Ly$\alpha$ clouds ($\log N(HI) <
14.13$) appears to be lower by 3--4 orders of magnitude than that in
the stronger, CIV absorbing clouds dealt with in paper II. Apparently,
the CIV gas has been affected more recently (and thus more
frequently, given it is unlikely that there is a sharply defined
"epoch" of star formation) by stellar feedback than the general
Ly$\alpha$ forest.

The fraction of the spectra that differ by more than 5\% in optical
depth with respect to the other line of sight, considering all
saturated regions as being disturbed, was measured for various column
density ranges.  The results imply a limit on the relative importance
of hydrodynamic disturbances in the presence of the restoring effects
of gas pressure. From the upper limit to the line of sight filling
factor ($f_{LOS} < 23\% $) we find upper limits to the {\it volume}
filling factor and space density of wind-producing galaxies as a
function of the mechanical wind luminosity, the density of the ambient
IGM, and the starting redshift of the wind.  The presence of short
lived, weak winds (low luminosity or high ambient density) is only
poorly constrained.  The other extreme, long-lived superwinds with
$L_{38}/n_{-5} \sim 1000$ arising as early as redshift 10, could  fill
up to 40\% of the volume of the universe by redshift 3.3, but not more
than 18\% if starting at redshift 4.  Currently there is no
observational evidence for the existence of $z\sim 10$ superwinds, but
closer to the epoch of our observations strong outflows have been seen
in high redshift galaxies (Franx et al. 1997; Pettini et al 2001), so
there is the possibility that a significant fraction of the volume of
the universe may have been stirred by winds; our results appear to show
that it is not a dominant one.

\acknowledgments
MR is grateful to NASA for supporting the initial phases of this
work under HF-01075.01-94A, and to the NSF for grant AST-0098492.
The work of WLWS was supported by NSF grant AST-9900733.
We thank Martin Haehnelt, Joop Schaye and an anonymous referee for helpful 
suggestions
and comments.

\pagebreak
 
\section{Appendix A}

Our criterion for the optimum choice of the maximum permissible
measurement error (chapter 3.2) is as follows:  we would like to omit $m$ of the $n$ measurements such that the
mean measurement error decreases from $\overline{\sigma_{\rm err}(n)}$
to $\overline{\sigma_{\rm err}(n-m)}$, as long as this reduced error 
outweighs the decrease in the number of data points available, and continues to lead to a reduction in the error of the determination of the total width of the
distribution to be measured, $\sigma_{tot}(n)/\sqrt{n} \rightarrow \sigma_{tot}(n-m)/\sqrt{n-m}$;
in other words, we throw away the data points with the largest errors
until the uncertainty in the width of the distribution does not improve
anymore. 
Data points should be continued to be omitted as long as
\begin{eqnarray}
E = \left(n-1 + S(m)\right)\ \overline{\sigma_{\rm err}^2(n)}
- \frac{n(n-1)}{n-m}\ \overline{\sigma_{\rm err}^2(n-m)} - \sigma_{\rm tot}^2(n)\ S(m) > 0,
\end{eqnarray}
where
\begin{eqnarray}
S(m)=\sum_{k=0}^{m-1} \frac{n(n-1)}{(n-k)(n-k-1)}.
\end{eqnarray}
For $m$ = 1, we obtain the simple relation
\begin{eqnarray}
\overline{\sigma_{\rm err}^2(n)} - \overline{\sigma_{\rm err}^2(n-1)} > \frac{\sigma_{\rm tot}^2(n)}{n}
\end{eqnarray}

The function $E$ is shown for the distributions of the velocity and column density differences in the bottom panels of figs. 5 and 8. The optimum value
of $E$ occurs where $E$ changes from positive to negative values.

\section{Appendix B}

This appendix addresses two questions:
(1) would a superbubble or wind blowing into the IGM at
high redshift produce detectable HI absorption, and
(2) can we produce limits on the volume filling factor of extragalactic
wind bubbles from the fraction of the Ly$\alpha$ forest spectrum 
disturbed by such events ?

\subsection{Detection of HI absorption from dense shells in the IGM}

We assume that  the dynamics of the wind shell can be described by the superbubble model
of Mac Low \& McCray (1988). Then the dependence of the radius $R$ on the 
mechanical luminosity $L = L_{38}\times 10^{38}$ erg s$^{-1}$, the time $t = t_7 \times 10^7$ years, and the number density of protons $n = n_{-5} 10^{-5}$ cm$^{-3}$ in the ambient IGM, is given by

\begin{eqnarray}
R = L_{38}^{1/5} n_{-5}^{-1/5} t_7^{3/5}\ 2670\ {\rm pc},
\end{eqnarray}
and the expansion velocity is
\begin{eqnarray}
\dot{R} = L_{38}^{1/5} n_{-5}^{-1/5} t_7^{-2/5}\ 157{\rm\ kms}^{-1}.
\end{eqnarray}
The {\t total} hydrogen column density seen radially outward through the shell is
\begin{eqnarray}
N_s(H) = n_s \Delta R = \frac{n R}{3} = 2.75\times 10^{16} L_{38}^{1/5} n^{4/5}_{-5} t_7^{3/5} {\rm cm}^{-2} \label{totcol}
\end{eqnarray}

The gas density $n_s$ in the shell is given by (Weaver et al 1977):
\begin{eqnarray}
n_s= n \frac{\dot{R}^2 + c^2}{c_{s}^2},
\end{eqnarray}
where $c$ is the sound speed in the undisturbed IGM 
and $c_s$ is the sound speed in the shell.
The thickness of the shell, $\Delta R$, is  
\begin{eqnarray}
\Delta R = \frac{N_s}{n_s} = \frac{n R}{3 n_s} = \frac{R c_s^2}{3(\dot{R}^2+c^2)}=
14.4 \frac{ L_{38}^{1/5} n^{-1/5}_{-5} t^{3/5}_7 c_{s,20}^2}
{ L_{38}^{2/5} n^{-2/5}_{-5} t^{-4/5}_7 + 0.0162 c_{20}^2} {\rm pc}.
\end{eqnarray}
The sound speeds are measured in units of 20 kms$^{-1}$.
The density in the shell is then
\begin{eqnarray}
n_s =  \frac{ 6.16\times 10^{-4}L_{38}^{2/5} n^{3/5}_{-5} t^{4/5}_7 + c_{20}^2}{c_{s,20}^2} {\rm cm}^{-3}.
\end{eqnarray}
It is not clear how exactly the gas is ionized. If it were in photoionization
equilibrium with an ionizing UV background radiation field with intensity
$J=10^{-21}$ erg cm$^{-2}$s$^{-1}$Hz$^{-1}$sr$^{-1}$ the 
neutral fraction would be approximately (Bergeron \& Stasinska 1986):
\begin{eqnarray}
\frac{n(HI)}{n(HII)} = 1.67\times10^{-3}\left(\frac{n_s}{10^{-2} {\rm cm}^{-3}}\right) \label{neutfrac}.
\end{eqnarray}
Thus the HI column density of the shell  (from equations \ref{totcol} and \ref{neutfrac}) is
\begin{eqnarray}
N_s(HI)=\frac{n(HI)}{n(HII)} N_s(H)= 2.8\times 10^{12}\ 
n_{-5}^{7/5} L_{38}^{3/5} t_7^{7/5} c_{s,20}^{-2} \left[1 + 0.0162 c_{20}^2 n_{-5}^{2/5} L_{38}^{-2/5} t_7^{-4/5}\right] {\rm cm}^{-2},
\end{eqnarray}
which is well detectable in absorption in our data.
The actual absorption is likely to be stronger (the bubble certainly sweeps up material
with higher than average density ($n_{-5}\sim 1$ at $z\sim 3$) in the vicinity
of the galaxy from which it originates. 

\subsection{Expanding shells as revealed by small scale differences in the
absorption pattern}

A dense shell crossing two adjacent lines of sight separated by a distance
$\Delta r$ will produce differences
in the column densities as long as the time it takes the shell to traverse  
the lines of sight  is shorter than the sound crossing time 
$\Delta r / c$ ($=5.8\times 10^6$ yr, in the present case).
From simple geometric arguments,
\begin{eqnarray}
\frac{\dot{R}}{R} b > c,
\end{eqnarray} 
where $b$ is the impact parameter of the lines of sight relative to the center
of the shell.
Thus the shell is detectable for $b > b_{min}= c R/\dot{R}$,
and the cross-section $\sigma$ of the shell for producing absorption lines with differences
between the lines of sight that have not been ironed out by pressure waves is
\begin{eqnarray}
\sigma &  = & \pi R^2 \left( 1 - \frac{c^2}{\dot{R}^2}\right),\ \   {\rm for}\ \   \dot{R}>c \\
       &  =  &    \ \ \ \ \ \ \ \ \ \ \    0,  \ \ \ \ \ \ \ \ \ \  {\rm for}\ \  \dot{R}\leq c.
\end{eqnarray}
The `footprint' of the shell absorption in the Ly$\alpha$ forest spectrum
is expected to consist of a pair of absorption lines, and we assume here
that each of these is broadened with a velocity width $\Delta  v_{fwhm} = 46$ kms$^{-1}$ (Doppler parameter 28 kms$^{-1}$,
which is typical for a photoionized Ly$\alpha$ line in the z=3 IGM). 
Conservatively, we count only the sum of the widths of the two
lines as contribution to the disturbance in the Ly$\alpha$ forest
spectrum as it is not clear whether the interior of the bubble would produce
detectable structure between the lines of sight (it is most likely to produce a clearing
in the forest without small scale features).

When expressed in spatial coordinates (at z=3.26) the combined velocity widths of the two lines
corresponds to a proper
distance $\Delta L = 2 \Delta v_{fwhm} H(z)^{-1} \approx 206 h_{50}^{-1}$kpc.
The fraction of the Ly$\alpha$ forest spectrum disturbed by a coeval population
of galaxies with space density $n_{gal}$, producing identical, spherically symmetric, expanding wind shells with radii
$R$, and expansion velocities $\dot{R}$ would be
\begin{eqnarray}
f_{LoS}  &=& n_{gal} \sigma \Delta L = 2 n_{gal} \pi R^2 \left( 1 - \frac{c^2}{\dot{R}^2}\right) \frac{\Delta v_{fwhm}}{H},\ \ \ \ {\rm for}\ \ \  \dot{R}>c,
\end{eqnarray}
whereas the {\it volume} filling factor for the wind bubbles is simply
\begin{eqnarray}
f_v = n_{gal} \frac{4\pi}{3}R^3 = \frac{2}{3}f_{LoS}R \left( 1 - \frac{c^2}{\dot{R}^2}\right)^{-1}\frac{H}{\Delta v_{fwhm}}.
\end{eqnarray}
Using the Mac Low \& McCray superbubble model to get a first crude estimate
the quantities $n_{gal}$ and $f_v$ can be expressed as functions of
the line of sight covering factor $f_{LoS}$, the time $t_7$ or redshift 
$z_{sf}$ when the wind (or star forming episode) started, and the ratio between the mechanical luminosity and the gas density in the IGM, $L_{38}/n_{-5}$.
The upper limits on $n_{gal}$ and $f_v$ corresponding to the upper limit
$f_{LoS}\leq 0.23$ are shown as a function of $z_{sf}$ for various values
of the parameter $L_{38}/n_{-5}$ in figures 11 and 12, respectively.
The transformation between time and redshift assumes $\Omega_m =0.3$ and $H_0$ = 50 km s$^{-1}$ Mpc$^{-1}$.

\pagebreak

\pagebreak

\clearpage

\section{figures}

\begin{figure}
    \centering
  \includegraphics[width=12.cm, angle=-90]{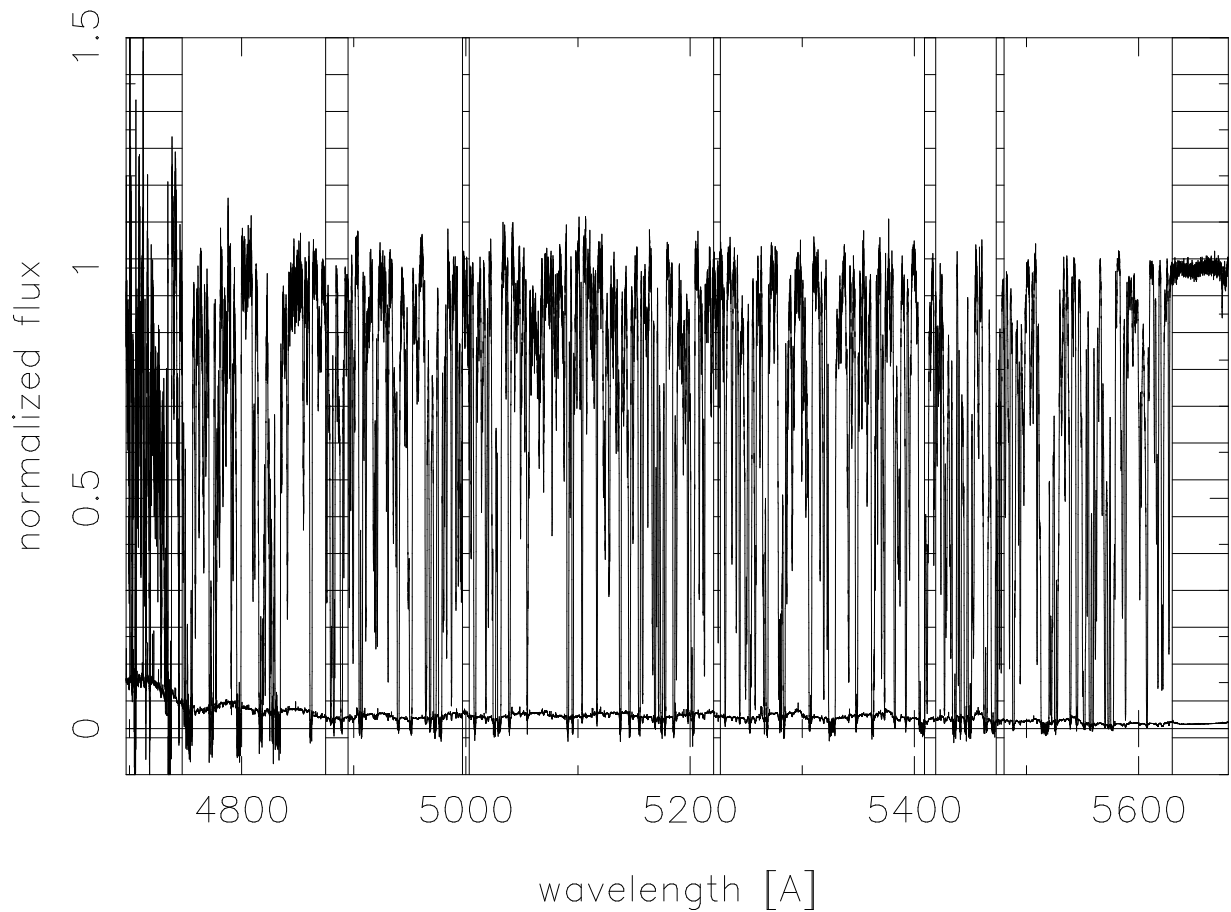}
         \caption{\small Normalized Ly$\alpha$ forest spectra of the QSO 1422+231 A image in the 
wavelength region between Ly$\beta$ and Ly$\alpha$ emission. The shaded
areas were excluded from the cross-correlation analysis as not belonging
to the forest or as being contaminated by known metal absorption lines.
The useable Ly$\alpha$ forest region extends over the wavelength range [4747, 5630] \AA\  or redshift range $z=\left[2.90, 3.63\right]$. The maximum beam separation at the short wavelength end of the spectrum is 270 $h_{50}^{-1}$\ {\em pc}. The slowly varying function at the bottom is the $1-\sigma$ error.
}
\end{figure} 

\begin{figure}
    \centering
  \includegraphics[width=11.cm, angle=-90]{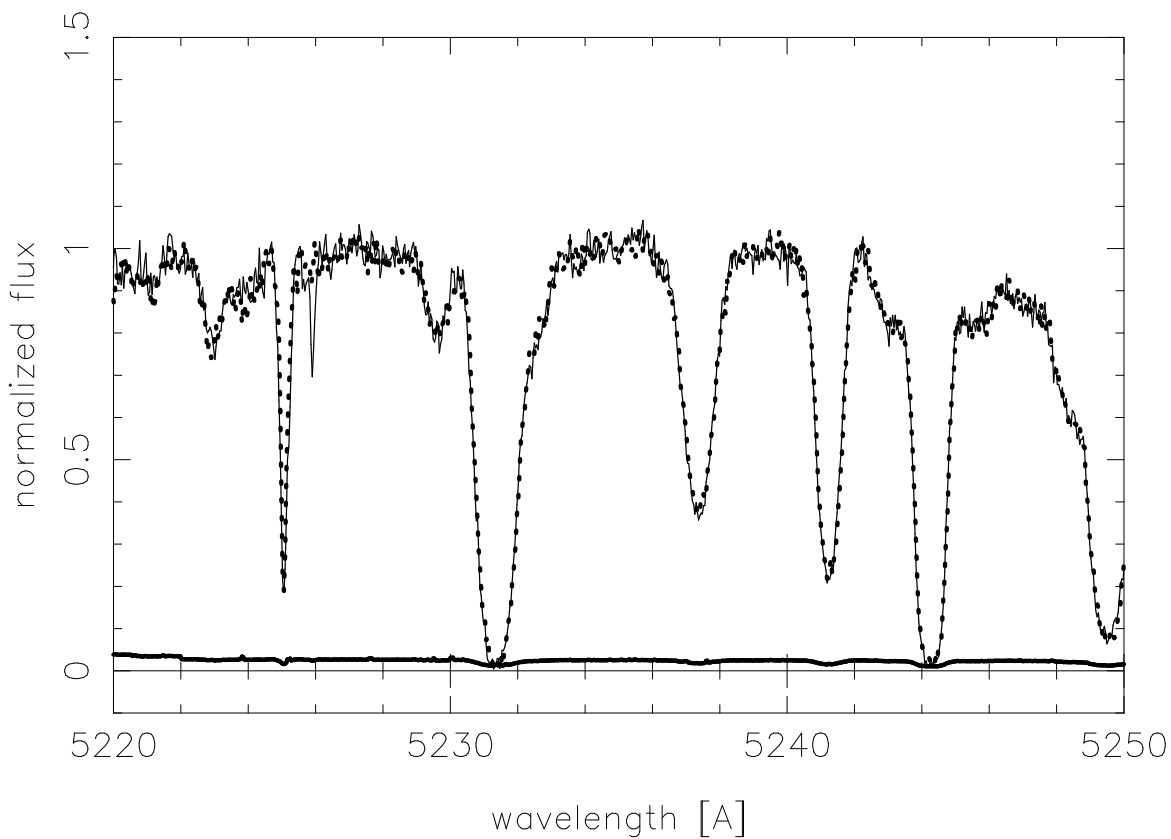}
         \caption{\small Enlargement of part of the spectrum of Q1422+231. The spectra of the A
(solid line) and C (dotted line) images are plotted on top of each
other.  There is hardly any difference between the spectra, with the
exception of a narrow line (which is largely absent in the spectrum of
the C image) and some lower column density fluctuations near 5226 \AA .
This line and the sharp stronger one at 5225 \AA\ can be identified
with a SiIV 1394\AA\ interloper from an absorption system at
$z=2.74889$. The other absorption features are plausibly attributed to
HI Ly$\alpha$.  The slowly varying function plotted at the bottom is the
$1-\sigma$ error of image A.  }
\end{figure} 

\begin{figure}
    \centering
  \includegraphics[width=11.cm, angle=-90]{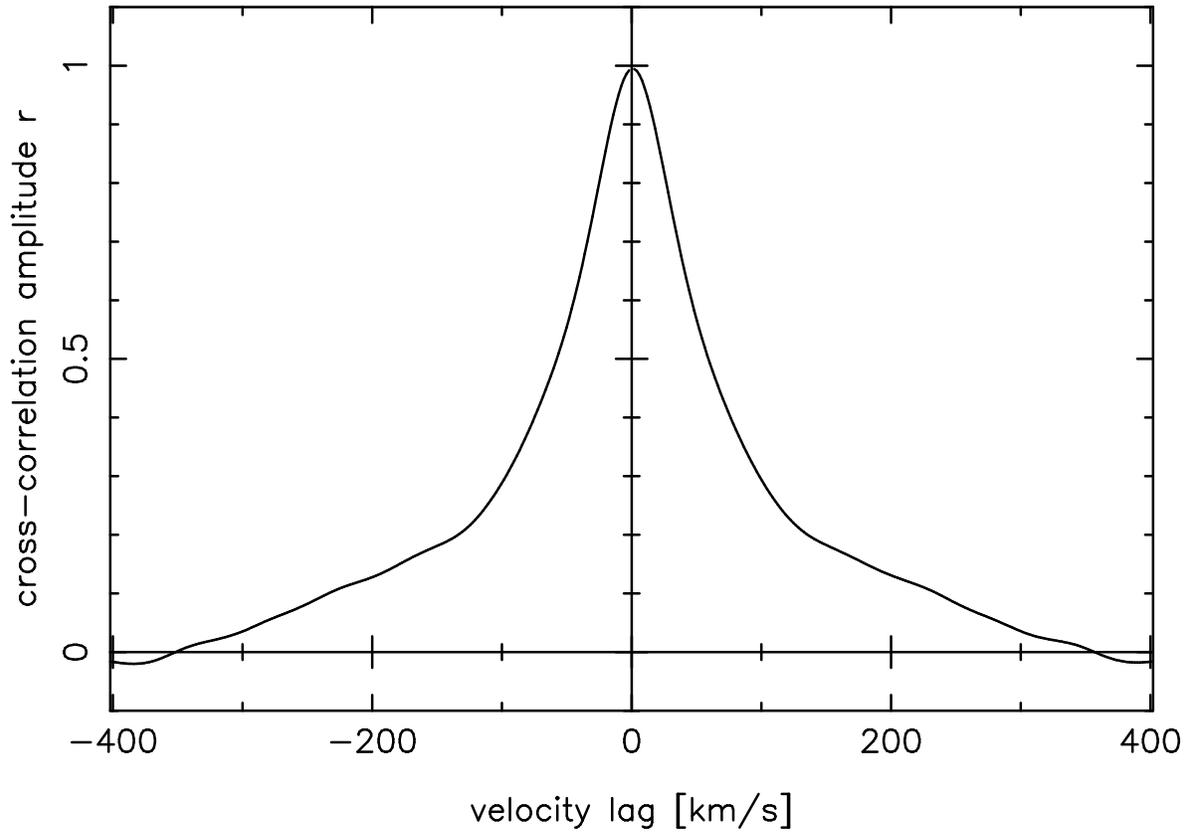}
         \caption{\small cross-correlation function $r(<d>,\Delta v$), for a mean beam
separation $<d>$ = 119\ $h_{50}^{-1}$\ {\em pc}, between the \op
forests in the Q1422+231 A and C image, for the spectral ranges shown
in figure 1. }
\end{figure} 

\begin{figure}
    \centering
  \includegraphics[width=11.cm, angle=-90]{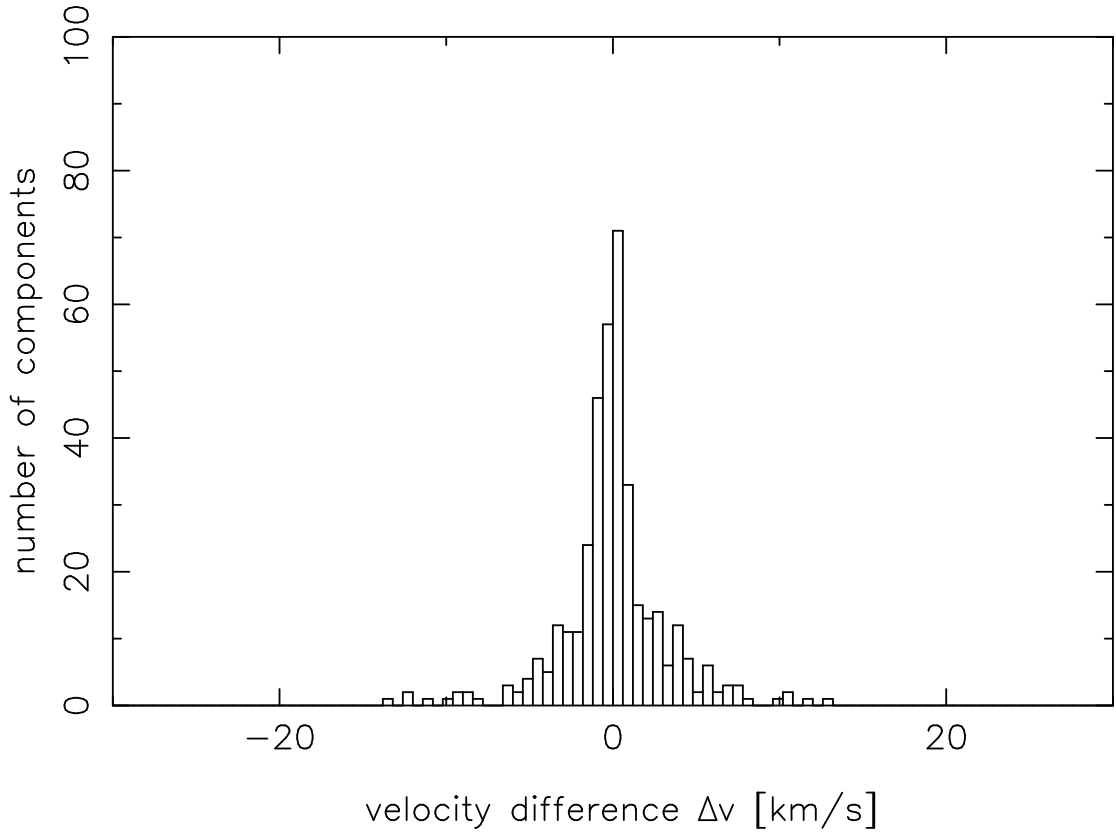}
         \caption{\small Distribution of the velocity differences $\Delta v$ (= $v_A$ -- $v_C$)
for pairs of absorption components. Two components were considered to
correspond to each other, when they were nearest neighbors, lay within
30kms$^{-1}$ of each other, and had column densities differing by less
than 50 \%.  }
\end{figure} 

\begin{figure}
    \centering
  \includegraphics[width=8.cm, angle=-90]{fig5a.ps}
  \includegraphics[width=8.cm, angle=-90]{fig5b.ps}
         \caption{\small Top: RMS difference between the
velocities of individually fitted absorption components in the two
lines of sight, versus the maximum measurement error permitted for the
redshifts in the sample used.  Solid dots give the observed width of
the $\Delta v$ distribution, open diamonds the width expected on the
basis of statistical errors in the velocity determination alone.  The
dashed shows the width of the observed distribution, convolved with a
Gaussian scatter with $\sigma = 1.5$ kms$^{-1}$. Bottom: function $E$
giving the  merit of increasingly getting rid off the data points with
the largest error (s.a. fig. 8).  The value of $E$ shows the trade-off
between omitting measurement points with errors above a certain value,
and the consequent reduction in the size of the sample, versus the
maximum measurement error accepted in the sample.  Positive values of E
imply that it is beneficial to continue throwing away points with the
largest errors, negative values imply a loss of information. The zero
crossing corresponds to the optimum choice. The meaning of $E$ is
explained in the text and the Appendix. }
\end{figure} 

\begin{figure}
    \centering
  \includegraphics[width=11.cm, angle=-90]{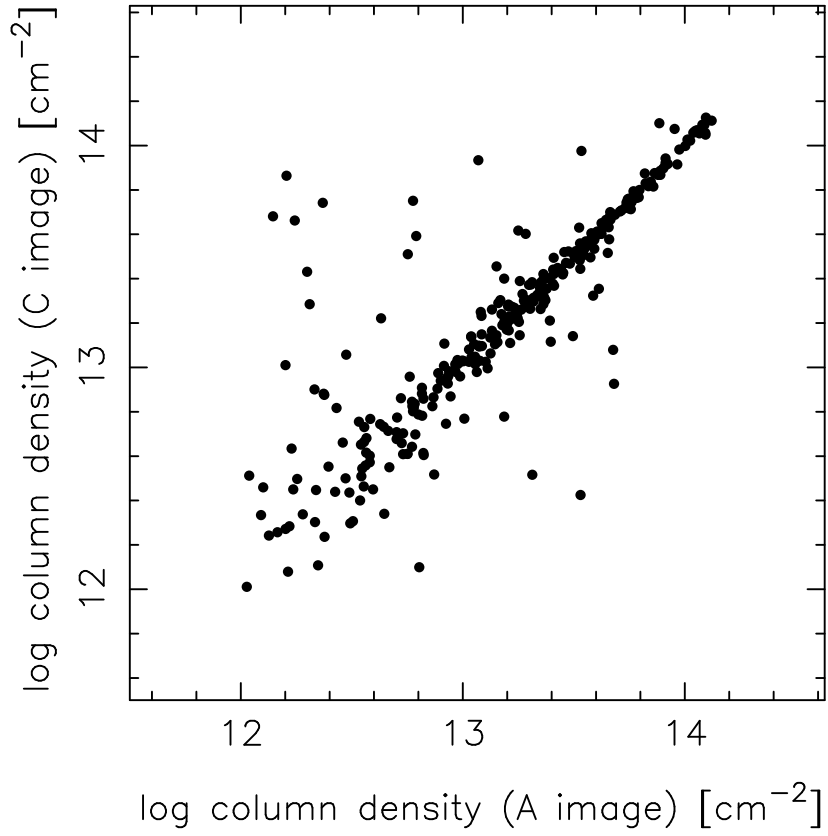}
         \caption{\small Scatter diagram of the logarithmic column densities in the A and C
spectrum. }
\end{figure} 

\begin{figure}
    \centering
  \includegraphics[width=11.cm, angle=-90]{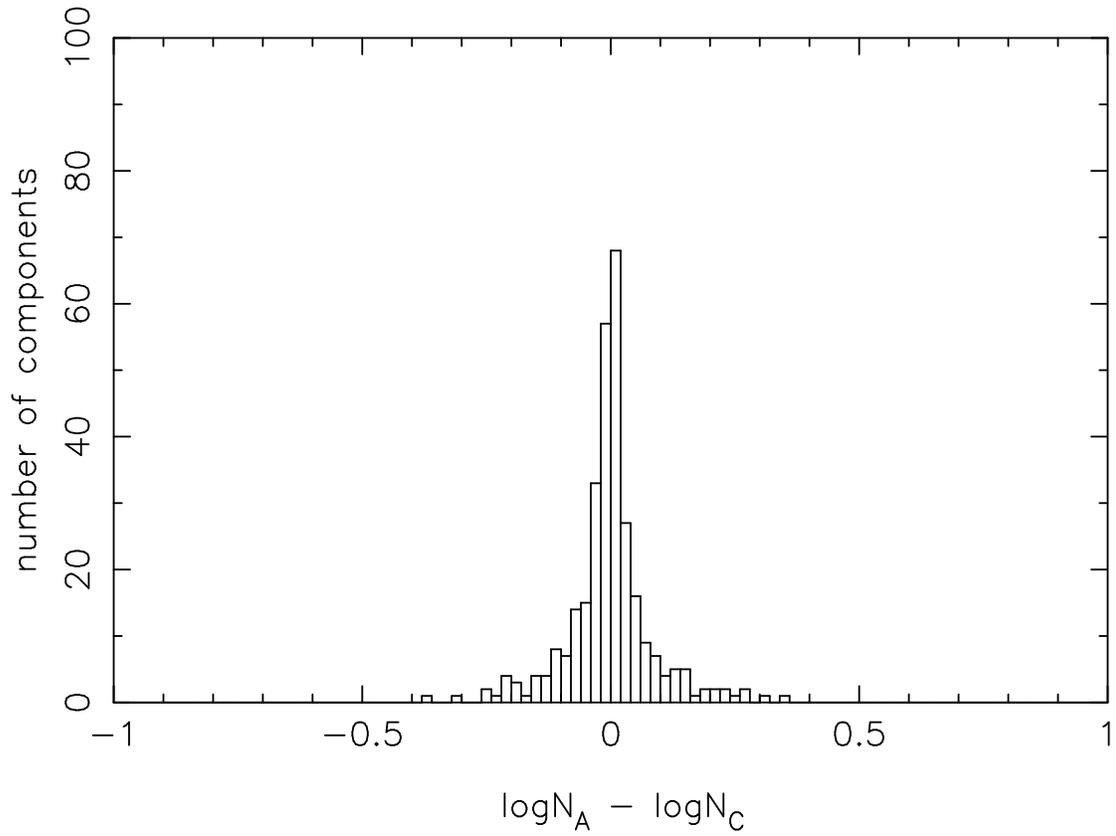}
         \caption{\small Distribution of the logarithmic column differences $\Delta \log N$ (=
$\log N_A$ -- $\log N_C$) for pairs of absorption components. Two
components were considered to correspond to each other when they were
nearest neighbors, lay within 30kms$^{-1}$ of each other, and had
logarithmic column densities differing by less than 1.0.  }
\end{figure}

\begin{figure}
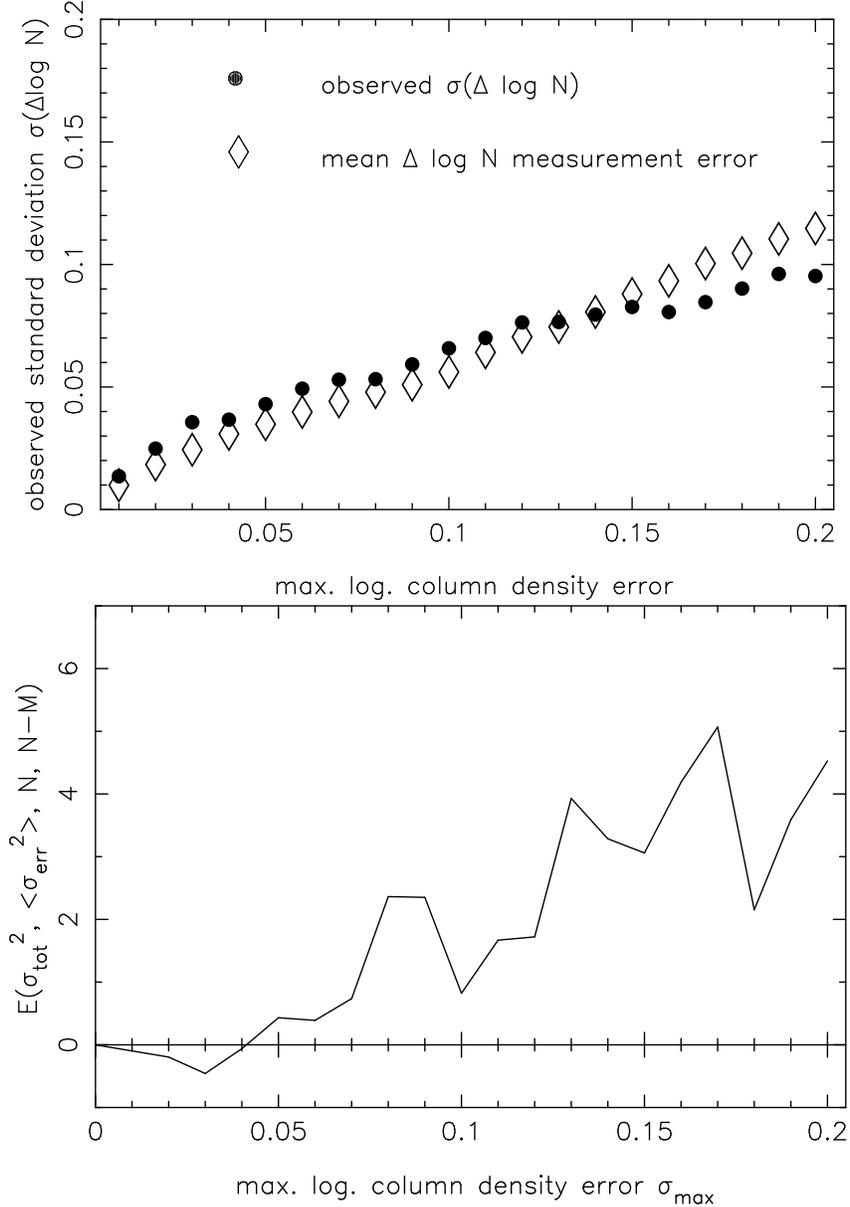

    \centering
  \includegraphics[width=8.cm, angle=-90]{fig8a.ps}
  \includegraphics[width=8.cm, angle=-90]{fig8b.ps}
         \caption{\small Top: RMS difference between the column densities of individually fitted
absorption components in the two lines of sight versus the maximum
measurement error permitted for the column densities in the sample used.
Solid dots give the observed width of the $\Delta\log N$ distribution,
open diamonds the width expected on the basis of statistical errors in
the column density determination alone. Bottom: the value of function $E$
shows the trade-off between omitting measurement points with errors
above a certain value, and the consequent reduction in the size of the
sample, versus the maximum measurement error accepted in the sample.
Positive values of E imply that it is beneficial to continue throwing
away points with the largest errors, negative values imply a loss of
information. The zero crossing corresponds to the optimum choice.
 }
\end{figure} 

\begin{figure}
    \centering
  \includegraphics[width=11.cm, angle=-90]{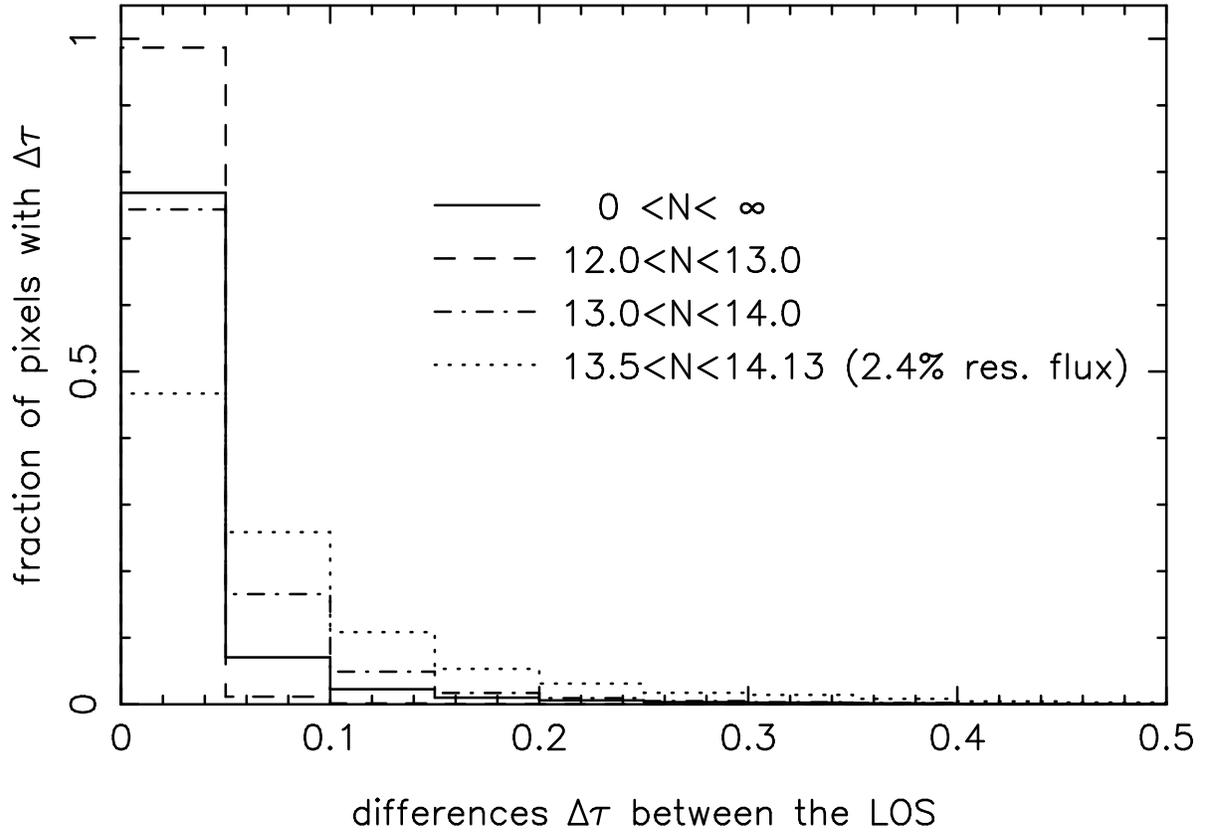}
         \caption{\small Distribution of differences between  the HI optical depth $\Delta\tau =
\mid\tau_A - \tau_C\mid$ for various column density ranges. The
fraction of pixels in the column density ranges shown are 100\% (solid
line), 36.6\% (dashed line), 36.4\% (dash-dotted line), and 17.4\%
(dotted line).   }
\end{figure} 

\begin{figure}
    \centering
  \includegraphics[width=12.cm, angle=-90]{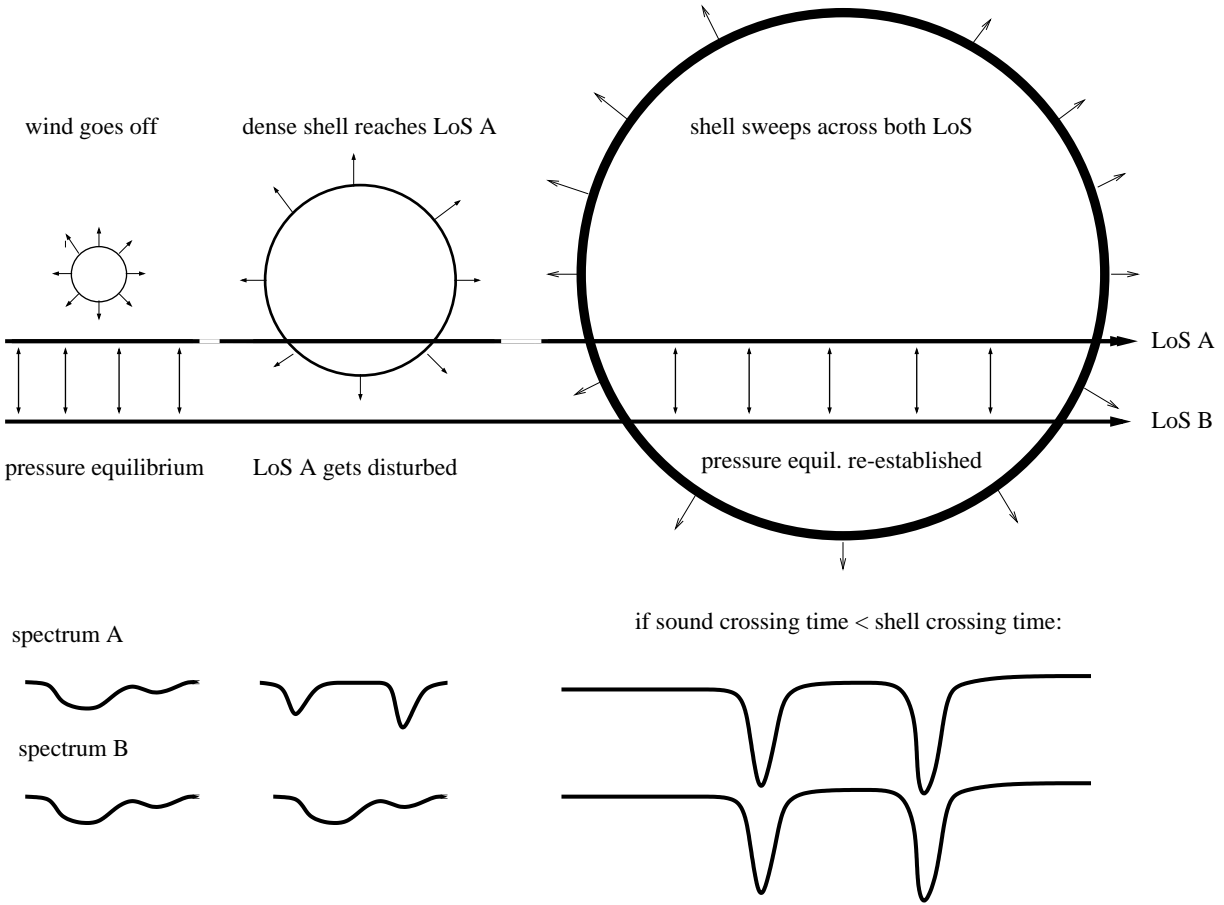}
         \caption{\small  left: A wind bubble expands near two lines of sight ($A$ and $B$). The
absorption spectra (bottom of the figure) look identical before the
shell hits the first line of sight. Middle: the first line of sight
(A) gets disturbed by the passage of the shell.  The absorption pattern of $A$
looks very different from that of $B$. Right:  after the shell has slowed
down sufficiently pressure waves re-establish equilibrium between the lines of
sight. The absorption pattern although different from the initial
state, looks again the same in both lines of sight.   }
\end{figure} 

\begin{figure}
    \centering
  \includegraphics[width=12.cm, angle=-90]{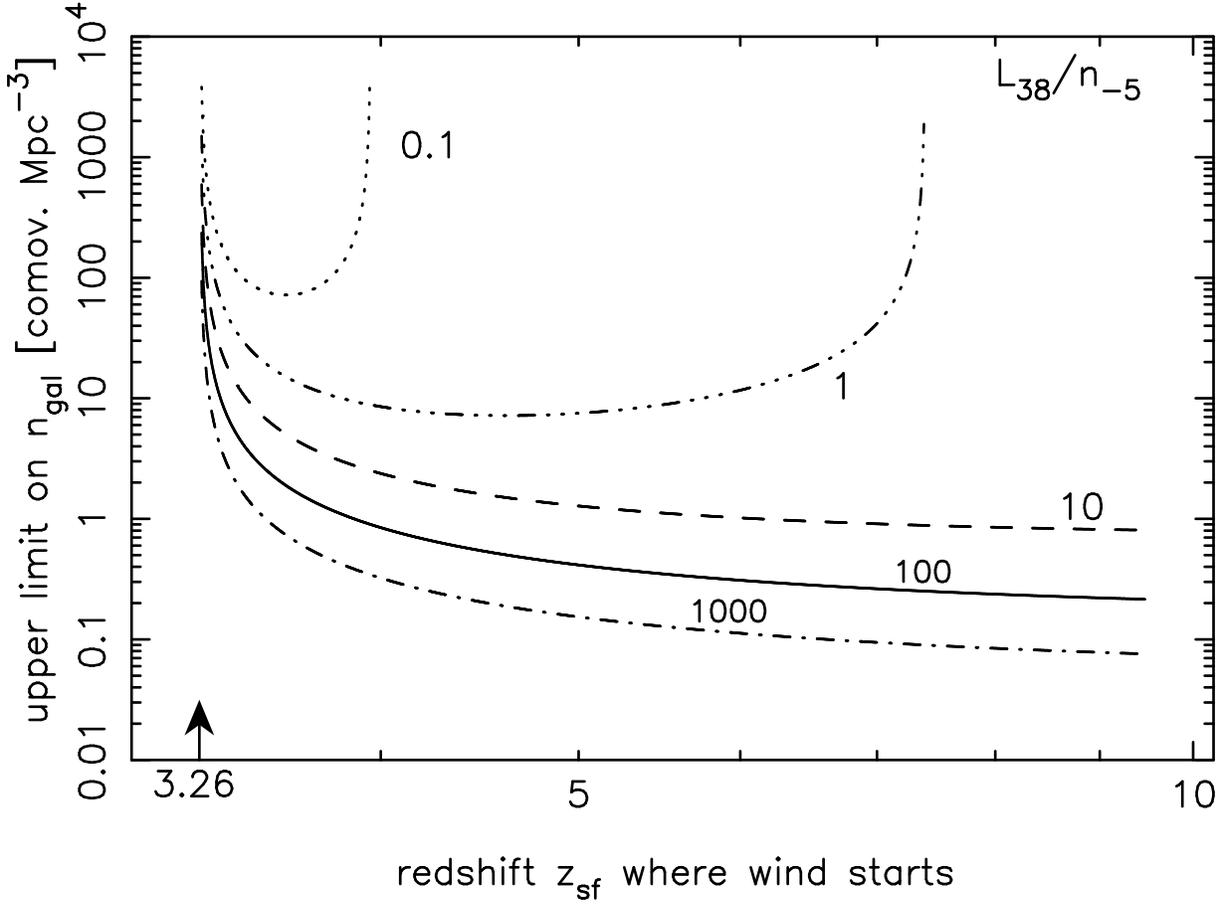}
         \caption{\small   Upper limit on the number density of galaxies surrounded by wind bubbles,
as a function of the redshift $z_{SF}$ where the wind starts.
The diagram is based on a `disturbed' fraction $f_{LoS}\leq 0.23$ of the Ly$\alpha$ forest. The different curves belong to different values of $L_{38}/n_{-5}$.
Winds at the low end of the  $L_{38}/n_{-5}$ range are less well constrained
because they have a shorter lifetime for detection. The mean redshift of observation (3.26) is marked by an arrow.   }
\end{figure} 

\begin{figure}
    \centering
  \includegraphics[width=12.cm, angle=-90]{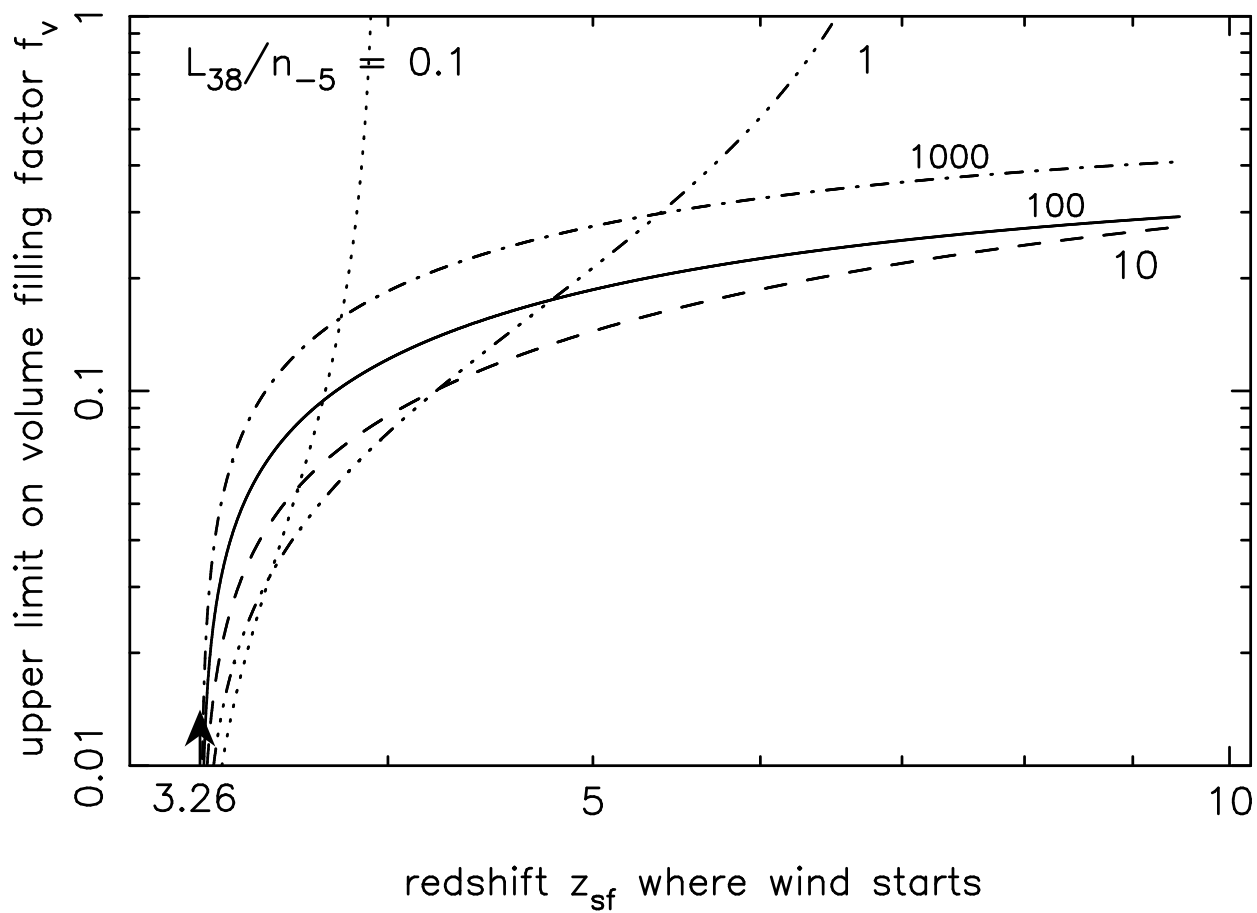}
         \caption{\small   Upper limit on the volume filling factor for wind blown bubbles. The
input parameters are the same as for the previous figure. Strong winds
$L_{38}/n_{-5} \sim 100-1000$ can fill a significant fraction of the
volume of the universe (up to 40\% if they start as early as redshift
10, up to $\sim 18\% $ if they start at z=4), whereas winds with $L_{38}/n_{-5} \leq 1$ are essentially
unconstrained if they occur before redshift 7.  }
\end{figure}

\end{document}